\documentclass[structabstract]{aa}  
\usepackage{graphicx}
\usepackage{txfonts}
\usepackage{psfig}
\usepackage{natbib}
\usepackage[english]{babel}  
\usepackage{url}

\begin{document}
   \title{XMM-Newton evidence of shocked ISM in SN 1006: indications of hadronic acceleration}

   \author{M. Miceli
          \inst{1,~2}
          \and
          F. Bocchino\inst{2}
	  \and
	  A. Decourchelle\inst{3}
	  \and
	  G. Maurin\inst{4}
	  \and
	  J. Vink\inst{5}
	  \and
	  S. Orlando\inst{2}
	  \and
	  F. Reale\inst{1,~2}
	  \and
	  S. Broersen\inst{5}
          }

   \institute{Dipartimento di Fisica, Universit\`a di Palermo, Piazza del Parlamento 1, 90134 Palermo, Italy\\
              \email{miceli@astropa.unipa.it}
         \and
INAF-Osservatorio Astronomico di Palermo, Piazza del Parlamento 1, 90134 Palermo, Italy
         \and
Service d'Astrophysique/IRFU/DSM, CEA Saclay, Gif-sur-Yvette, France
	 \and
Universit\'e de Savoie, 27 rue Marcoz, BP 1107 73011-Chambery cedex, France
\and
Astronomical Institute ``Anton Pannekoek", University of Amsterdam, P.O. Box 94249, 1090 GE Amsterdam, The Netherlands	
             }

\date{}

 
  \abstract
   {Shock fronts in young supernova remnants are the best candidates for being sites of cosmic ray acceleration up to a few PeV, though conclusive experimental evidence is still lacking.}
   {Hadron acceleration is expected to increase the shock compression ratio, providing higher postshock densities, but X-ray emission from shocked ambient medium has not firmly been detected yet in remnants where particle acceleration is at work. We exploited the deep observations of the XMM-Newton Large Program on SN 1006 to verify this prediction.}
   {We performed spatially resolved spectral analysis of a set of regions covering the southeastern rim of SN~1006. We studied the spatial distribution of the thermodynamic properties of the ambient medium and carefully verified the robustness of the result with respect to the analysis method.}
   {We detected the contribution of the shocked ambient medium. We also found that the postshock density of the interstellar medium significantly increases in regions where particle acceleration is efficient. Under the assumption of uniform preshock density, we found that the shock compression ratio reaches a value of $\sim6$ in regions near the nonthermal limbs.}
   {Our results support the predictions of shock modification theory and indicate that effects of acceleration of cosmic ray hadrons on the postshock plasma can be observed in supernova remnants.}


   \maketitle
%

\section{Introduction}

Supernova remnant (SNRs) are strong candidates for being the main source of energetic cosmic rays up to at least $3\times10^{15}$ eV (\citealt{be87,bv07}).
X-ray synchrotron emission from high-energy electrons accelerated at the shock front up to TeV energies was first observed in SN~1006 \citep{kpg95} and then in other young SNRs \citep{rey08}, but a firm detection of high-energy hadrons is still lacking. 

Recently, TeV emission has been detected in SN~1006 \citep{aaa10} and in a handful of SNRs showing bright nonthermal X-ray emission (\citealt{aab07a,aab07,aaa07m,aaa09,aaa11}). The origin of the gamma-ray emission can be leptonic (i.~e., inverse Compton from the accelerated electrons) and$/$or hadronic (i.~e., proton-proton interactions with $\pi^0$ production and subsequent decay).
The hadronic scenario would directly prove that SNRs can accelerate cosmic rays up to PeV energies. Unfortunately, it is not easy to unambiguously ascertain the origin of the gamma-ray emission, as, for example, in RXJ1713.7-3946, where both hadronic \citep{bv10} and leptonic \citep{aaaa11,eps10,psr10} scenarios have been invoked to explain the observed multiwavelength observations. 
However, different indications concur in supporting the presence of high-energy hadrons in some young SNRs (e.~g., Tycho, \citealt{ekh11,mc12}, and RCW 86, \citealt{hvb09}).
As for SN~1006, a pure leptonic model is consistent with the observations, but a mixed scenario that includes leptonic and hadronic components also provides a good fit to the gamma-ray data \citep{aaa10}.

An alternative way to reveal hadron acceleration in SNRs is to probe its effects on the shock dynamics. The nonlinear back-reaction of high-energy particles on background plasma is predicted to strongly modify the shock properties by increasing the shock compression ratio and decreasing the postshock temperature with respect to the expected Rankine-Hugoniot values \citep{be99,deb00,bla02,vyh10}. This effect is known as ``shock modification". The observational test of these predictions requires accurate diagnostics of the thermal X-ray emission from the shocked interstellar medium (ISM). However, the ISM contribution in the X-ray spectra of remnants where particle acceleration is efficient is typically masked out by the bright synchrotron emission and by the thermal emission from shocked ejecta, as in RXJ1713.7-3946 \citep{abd09},  Vela Jr \citep{paf10}, G1.9+0.3 \citep{brg10}, and Tycho \citep{chb07}, and has not been firmly detected so far.
In SN~1006 the quest for X-ray emission from shocked ISM is also ongoing \citep{abd07,mbi09}.

SN~1006 exhibits a morphology characterized by two opposed radio, X-ray, (and $\gamma-$ray) bright limbs dominated by nonthermal emission (bilateral SNR) and separated by an inner region of low surface brightness and soft, thermal X-ray emission. 
Thermal X-ray emission has been associated with shocked ejecta \citep{abd07,mbi09}, consistent with the detection of Fe-rich plasma in the interior of the shell \citep{ykk08}. A model with a nonthermal component plus a thermal component associated with the ejecta can fit the archive \emph{XMM-Newton} spectra. An additional thermal ISM component is not needed from a statistical point of view \citep{abd07,mbi09}, though some hints about ISM temperature and density ($kT_{\rm ISM}\sim 1.5-2$ keV, $n<0.2$ cm$^{-3}$) can be inferred \citep{abd07}.
\emph{Suzaku} spectra of SN~1006 have been modeled with three thermal components only by assuming that the soft component originated in the ISM; nevertheless, it was not possible to exclude that the O lines, which dominate the soft emission, originate in the ejecta \citep{ykk08}. A firm detection of X-ray emission from the ISM is therefore still lacking.

Indirect evidence for shock modification in SN~1006 has been obtained by measuring the distance, $D_{SFCD}$, between the shock front and the contact discontinuity, which is expected to be smaller in nonthermal limbs where particle acceleration is more efficient. $D_{SFCD}$ is instead almost the same all over the shell (even in regions dominated by thermal emission), though it is much smaller than that expected from unmodified shocks \citep{chr08,mbi09}. Recently, 3-D magneto-hydrodynamic simulations have shown that this small distance can be naturally explained by ejecta clumping, without invoking shock modification \citep{obm12}. Therefore, the small value of $D_{SFCD}$ is not a reliable indicator of hadronic acceleration. 

Here new, deep observations of the \emph{XMM-Newton} SN~1006 Large Program allow us to present the detection of X-ray emission from shocked ISM in SN~1006 and to show indications for the effects of hadron acceleration on the postshock density of the ambient medium. 

The paper is organized as follows: in Sect. \ref{DP}, we describe the data analysis procedure and the background subtraction; in Sect. \ref{Results}, we show the results of the spatially resolved spectral analysis; and, finally, we discuss our conclusions in Sect. \ref{Conclusions}.


\section{Data analysis and background subtraction}
\label{DP}

In this work, we analyze the data obtained within the \emph{XMM-Newton} Large Program of observations of SN~1006 (PI A. Decourchelle, 700 ks of total exposure time). These observations were all performed with the medium filter by using the full frame mode for the MOS cameras and the extended full frame mode for the pn camera, and were processed with the Science Analysis System. Light curves, images, and spectra were created by selecting events with PATTERN$\le$12 for the MOS cameras, PATTERN$\le$4 for the pn camera, and FLAG=0 for both. To reduce the contamination by soft proton flares, the original event files were screened by using the sigma-clipping algorithm (ESPFILT tasks). The images were produced by adopting the procedure described in \citet{mbi09} and are all background-subtracted, vignetting-corrected, and adaptively smoothed to a signal-to-noise ratio of ten.

Spectral analysis was performed in the energy band $0.3-7$ keV using XSPEC. The source regions $a-g$ (see Sect. \ref{Results} and Fig. \ref{fig:1006}) analyzed in this paper are covered by the observations ID 0555630101 and 0555631001, while region $h$ is covered also by observation ID 0653860101. The screened MOS1$/$MOS2$/$pn exposure times are $43/42/26$ ks, $59/60/43$ ks, and $101/104/71$ ks for observations 0555630101, 0555631001, and 0653860101, respectively. MOS and pn spectra of the different observations were fitted simultaneously. Pn spectra were analyzed only in the $0.6-3$ keV band to reduce possible contamination from the instrumental background (that can be high above 3 keV) and cross-calibration issues (below 0.6 keV). The results of the fittings of the pn spectra in the 0.6-7 keV band are shown in the Appendix.

The spectral model adopted here is the same as that in \citet{mbi09} (VPSHOCK component for the ejecta plus nonthermal emission) with the addition of a VPSHOCK component with solar abundances to describe the ISM. VPSHOCK model \citep{blr01} describes an isothermal and optically thin plasma in nonequilibrium of ionization with a linear distribution of ionization timescale versus emission measure (its parameters are electron temperature, plasma emission measure, and ionization parameter). The nonthermal emission is described by the SRCUT \citep{rk99} model (synchrotron emission from an electron power law with exponential cut-off). In the SRCUT component, we fixed the radio-to-X-ray photon index to $\alpha=0.5$ \citep{ahs08,mbi09}, while the normalization was derived from the radio image \citep{mbi09}. We verified that the results do not change significantly by letting $\alpha$ free to vary in the fittings, though error bars grow higher. In the ejecta component, the S abundance is the same as the Si abundance.
 We  included three narrow Gaussians in our model to take account of the $K_\delta$, $K_\epsilon$, and $K_\zeta$ O~VII lines that are not included in the spectral code. We also added a systematic 2\% error term to reflect the estimated uncertainties in the calibration of the instrumental effective area (see the EPIC status of calibration report XMM-SOC-CAL-TN-0018).  The interstellar absorption is described by the TBABS model, and the absorbing column is set to $N_H=7\times10^{20}$ cm$^{-2}$, in agreement with \citet{dgg02}. We verified that the best-fit values (and particularly the ISM emission measure) do not change significantly by letting the $N_H$ vary freely in the fitting procedure (though the uncertainties in the best-fit parameters become higher). 
 

To check the validity and the robustness of our result, we adopted three different procedures for background subtraction:
\begin{enumerate}
\item The best-fit values presented in Sect. \ref{Results} and discussed in Sect. \ref{Conclusions} were obtained by subtracting a background spectrum extracted from a nearby region immediately outside of the SNR shell, $BG_{tot}$, from the spectrum extracted from our source region, $SP$ (source $+$ background). By extracting $BG_{tot}$ from different regions outside of the shell, we verified that the best-fit values do not depend on the choice of the background region since they do not vary significantly. 
\item We also adopted a ``double background-subtraction'' technique: we considered the instrumental background, $BG_i$ (extracted from the same region in detector coordinates as $SP$) in filter-wheel closed data and computed $SP-(BG_{tot}-\beta~BG_i)-\beta~BG_i$, where $\beta$ is the ratio of the flux from the unexposed corners of archived observations to that of our SN~1006 observations.
\item We produced the response files for the $BG_{tot}$ spectrum. We then modeled the ``background'' spectra in XSPEC and fixed these background parameters when fitting the $SP$ spectra. 
\end{enumerate}
 We verified that all the procedures provide consistent results. In particular, the values of the ISM temperature obtained with the different methods are all consistent within two sigmas, while those of the ISM emission measure are all consistent at less than one sigma\footnote{The different subtraction methods slightly affect the high-energy tail of the spectrum, and small variations of the signal at high energy are modeled in the fittings with a higher$/$lower temperature of the ISM component (and a higher$/$lower cut-off frequency for the SRCUT component).}.  

We used the best-fit values of temperature and emission measure to derive the ISM and ejecta densities shown in Sect. \ref{Results} and the filling factor $(f_{\rm ISM},~f_{ejecta})$ of each component, obtained as in \citet{bms99}. These quantities were derived by assuming a pressure equilibrium between ISM and ejecta and by calculating the volume of the X-ray emitting plasma (according to the formula presented in the Appendix).

\section{Results}
\label{Results}

The bilateral morphology of SN~1006 suggests regions at the rim with highly efficient particle acceleration (the nonthermal limbs), separated by regions with less efficient particle acceleration. We therefore expect to observe stronger shock modification near the nonthermal limbs than in the thermal regions. 

\begin{figure}[htb!]
 \centerline{\hbox{     
     \psfig{figure=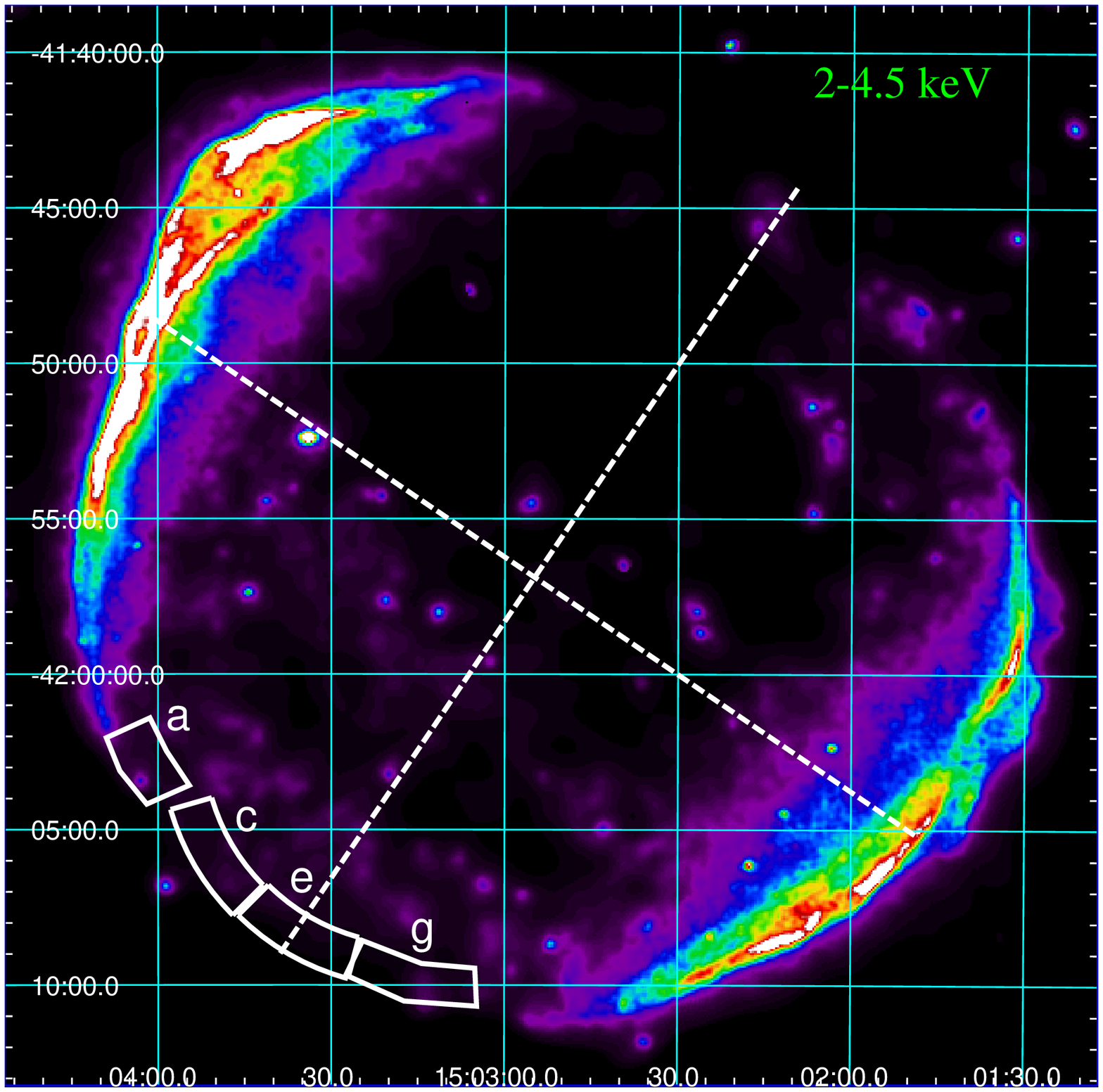,width=\columnwidth}}}
      \centerline{\hbox{     
     \psfig{figure=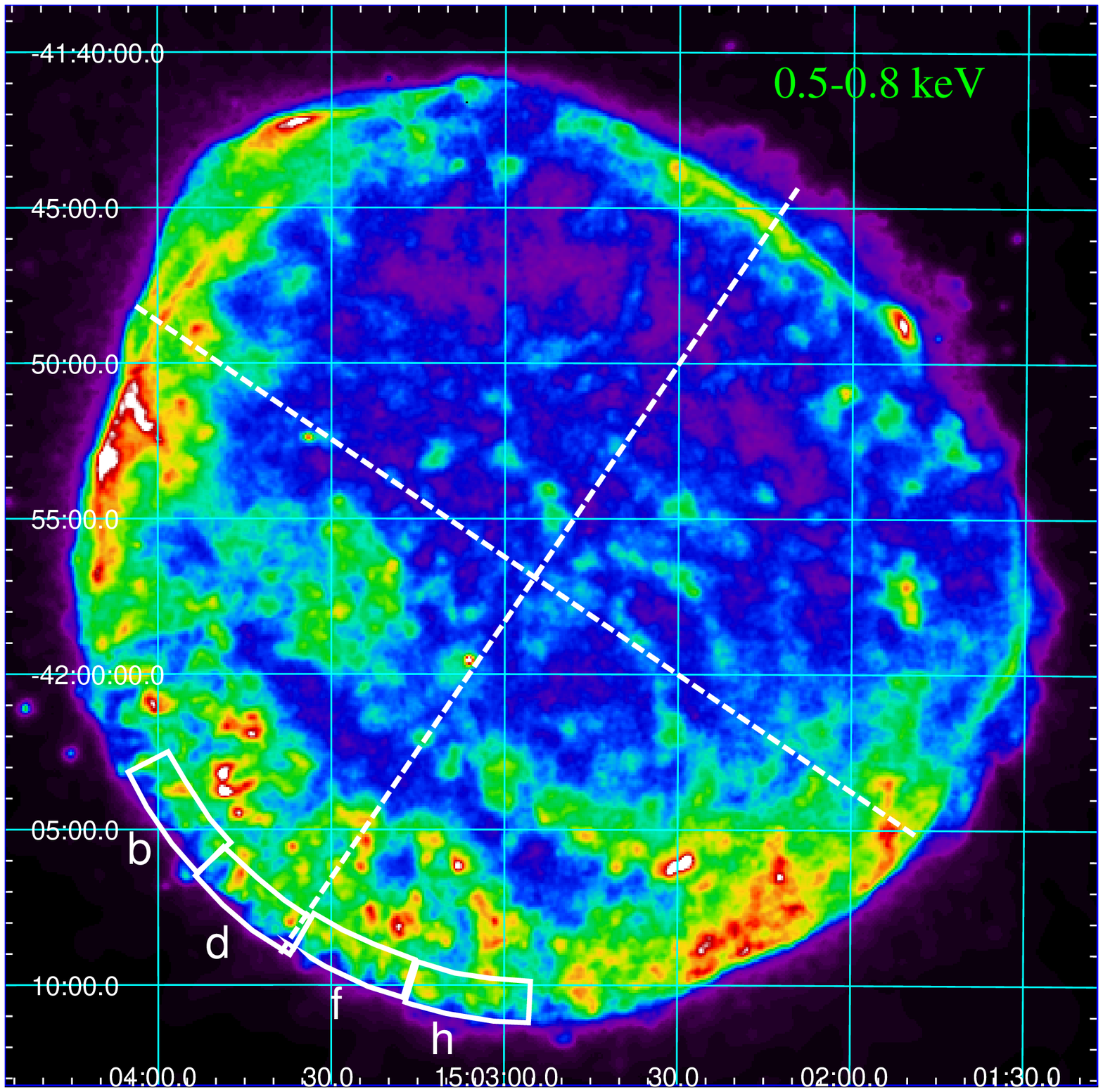,width=\columnwidth}}}  
 
\caption{\emph{Upper panel}: mosaicked count-rate images (MOS-equivalent counts per second per bin) of SN~1006 in the $2-4.5$ keV band. The bin size is $4''$ and the image is adaptively smoothed to a signal-to-noise ratio ten. The regions selected for the spectral analysis of the rim are superimposed (four in the upper panel and four in the lower panel for clarity). North is up and East is to the left. The two dashed lines indicate the symmetry axis of the remnant, marking the center of the synchrotron limbs (northeast and southwest) and of thermal limbs (northwest and southeast). \emph{Lower panel:} same as upper panel in the $0.5-0.8$ keV energy band.}
\label{fig:1006}
\end{figure}

Figure \ref{fig:1006} shows an image of SN~1006 in the $2-4.5$ keV band (upper panel), where the X-ray emission is mainly nonthermal, and in the $0.5-0.8$ keV band (lower panel), where thermal emission dominates. The figures also shows the symmetry axis of the remnant and the regions selected for the spatially resolved spectral analysis. The regions cover the southeastern thermal limb of SN~1006, where the preshock density of the ISM is expected to be fairly uniform (as indicated by the circular shape of the shock front and in agreement with \citealt{dgg02}). We first focus on the middle of the southeastern thermal rim (region $e$ in Fig. \ref{fig:1006}), where the contribution of the synchrotron emission is the smallest and where we expect minimum shock modification. 

\subsection{Detection of the ISM component}

\begin{figure}[t!]
 \centerline{\hbox{     
     \psfig{figure=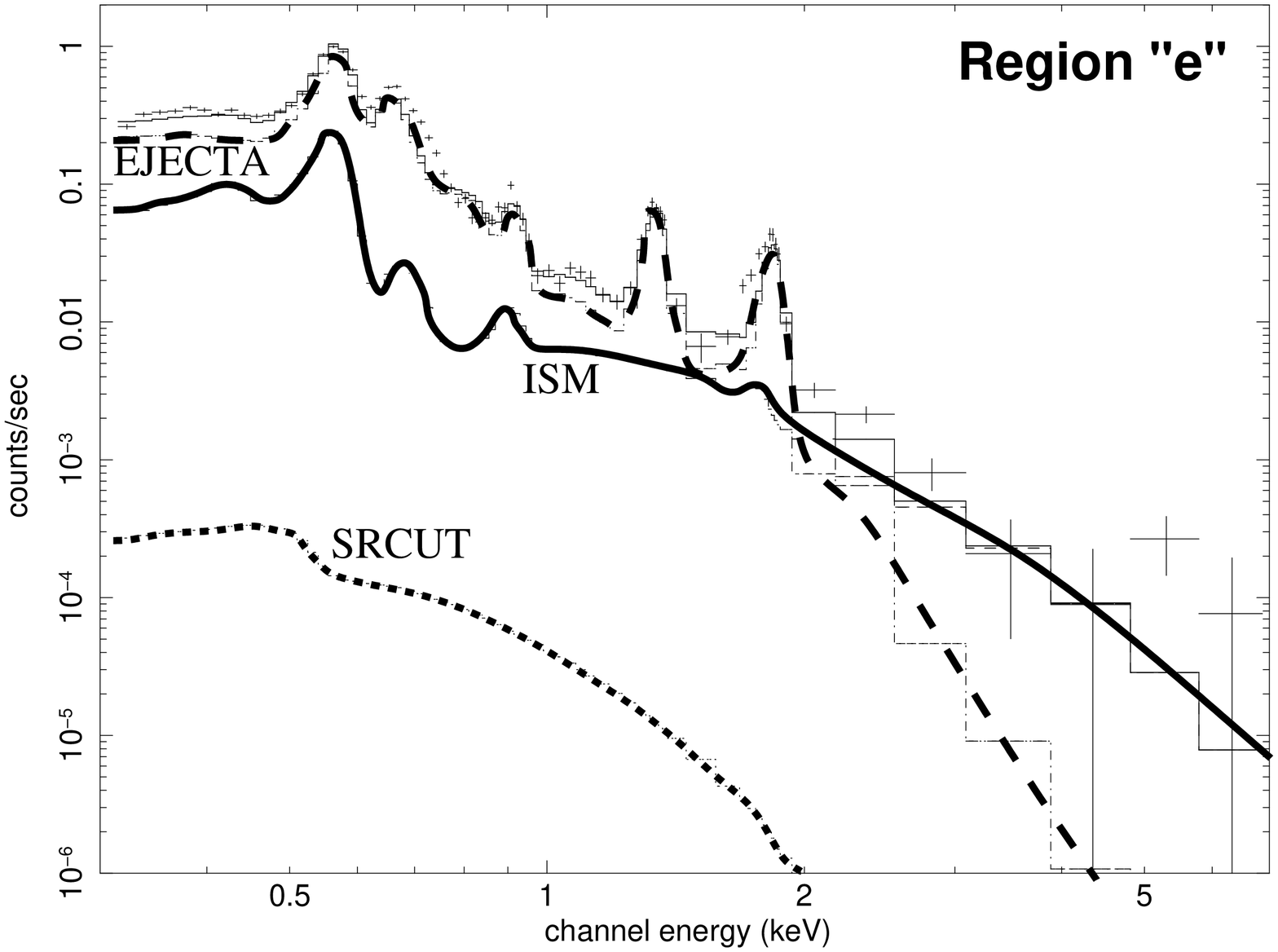,width=\columnwidth}}}
      \centerline{\hbox{     
     \psfig{figure=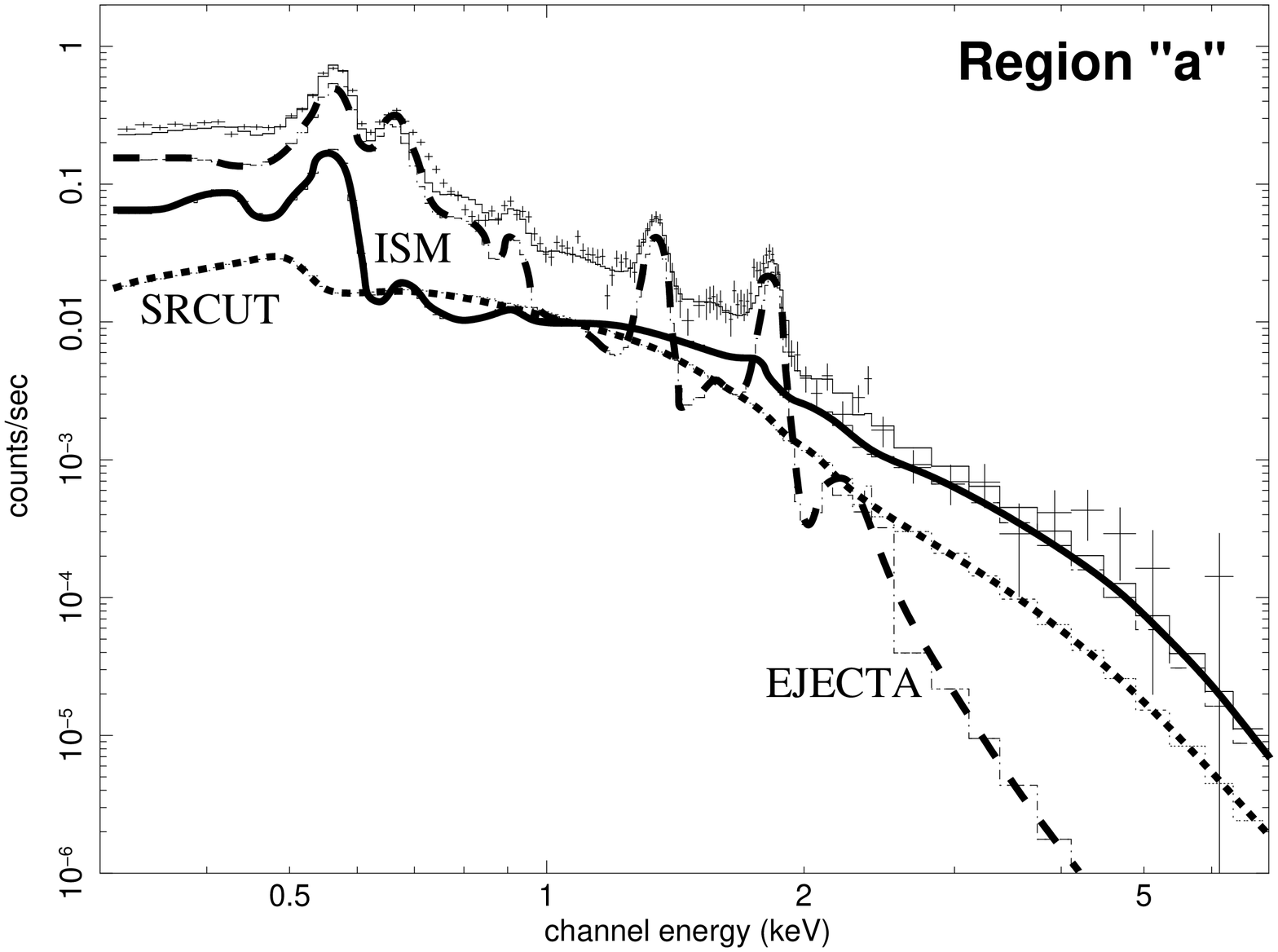,width=\columnwidth}}}  
 
\caption{\emph{Upper panel}: Spectra of region $e$ (see Fig. \ref{fig:1006}) with the corresponding best-fit model (see text). The contribution of each thermal (ISM, solid line, and ejecta, dashed line) and nonthermal (SRCUT, dotted line) components is shown. Only MOS spectra are shown for clarity. \emph{Lower panel}: same as upper panel for region $a$.}
\label{fig:spec}
\end{figure}

Using the deep observations of the Large Program, we found that a model with three components (ejecta, ISM, and synchrotron emission) fits the spectrum of region $e$ significantly better than a model with only two components. 
A thermal component for the ejecta (optically thin plasma in nonequilibrium of ionization with free O, Ne, Mg, and Si abundances, plus the nonthermal SRCUT component, yields a $\chi^2=1199.8$ (with $658$ d.~o~.f.). By adding another VPSHOCK component with solar abundances, we get a lower value ($\chi^2=1112.0$ with $655$ d.~o~.f.). 
The additional component is significant at $>11$ sigmas. We carefully verified the robustness of this result with respect to the analysis method and the background subtraction procedure, as explained in Sect. \ref{DP}. 
The upper panel of Fig. \ref{fig:spec} shows the spectrum of region $e$ with its best-fit model.

The origin of the additional thermal component can be attributed to shocked ISM. This is supported by the values of the parameters derived from the spectral analysis: i) the postshock density $n_{\rm ISM}=0.14\pm0.01$ cm$^{-3}$ indicates a preshock density $n_0<0.05$ cm$^{-3}$, consistent with the high galactic latitude of the remnant and in agreement with the upper limit derived from previous X-ray data \citep{abd07}; ii) the ionization parameter (i.~e., the integral of density over time since the passage of the shock front) $\tau_{\rm ISM}\sim7\times10^8$ s cm$^{-3}$ provides a very reasonable estimate of the time elapsed after the shock impact ($\sim 200$ yr), considering the 1 kyr age of the remnant; and iii) the electron temperature ($\sim 1.4$ keV) suits the expectations for the shocked ISM, as shown below. In collisionless shocks at high Mach number (shock speed of the order of $10^3$ km/s), the electron-to-proton temperature ratio is expected to be $T_e/T_p\ll 1$ \citep{glr07,vlg03}. In the 
northwestern limb of SN~1006, the shock velocity is $v_{sh}\sim3000$ km$/$s and it has been calculated that $T_e/T_p<0.07$ \citep{gwr02}. Assuming the same ratio (and no or little shock modification) in the southeastern region, where the shock speed is $\sim5000$ km$/$s \citep{kpl09},  we get $kT_e<3.5$ keV, in agreement with the temperature above. Moreover, a model developed for RXJ1713.7-3946 with parameters similar to those of SN~1006 (but different mass of ejecta) and including the effects of particle acceleration and electron heating by Coulomb collisions with shocked protons has shown that $kT_e$ is expected to rapidly reach 1 keV before leveling out between one and two keV \citep{eps10}, in remarkably good agreement with our results. Similar values of the electron-to-proton temperature ratio have also been obtained by modeling the Tycho SNR (\citealt{mc12}).
We therefore conclude that the additional thermal component originates in the shocked ISM.

\begin{figure}[b!]
 \centerline{\hbox{     
     \psfig{figure=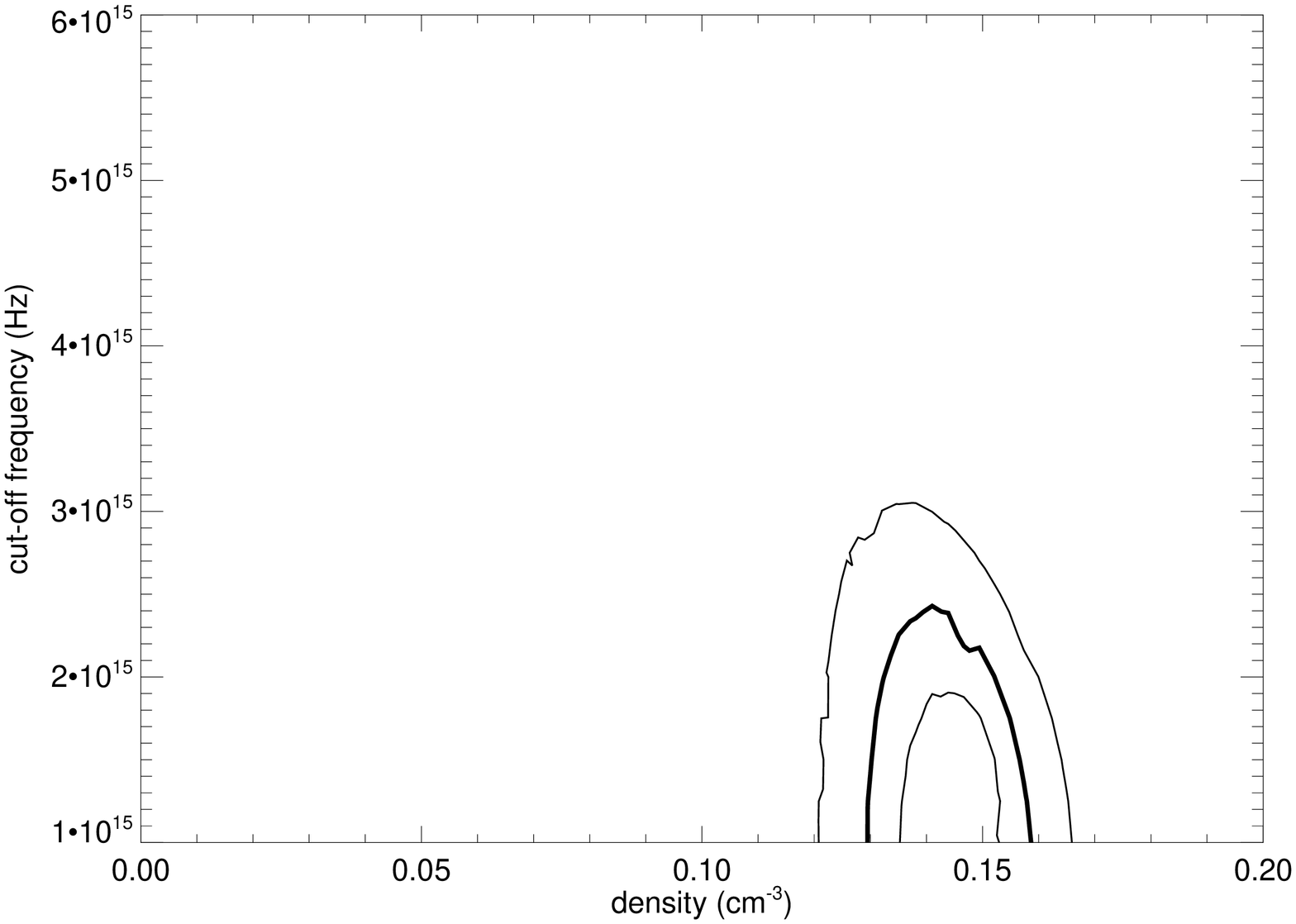,angle=90,width=\columnwidth}}}
  \centerline{\hbox{     
     \psfig{figure=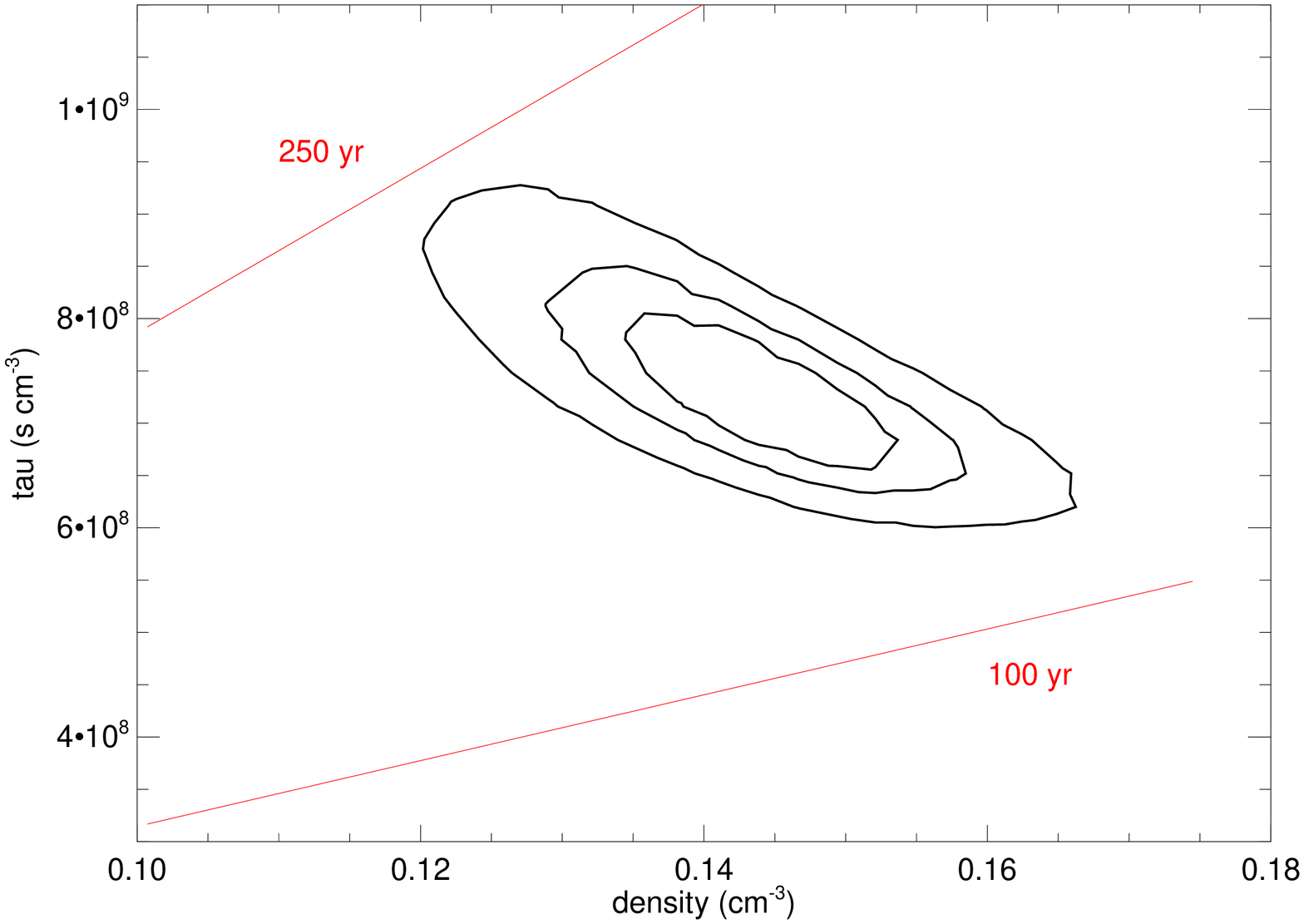,angle=90,width=\columnwidth}}}    
\caption{\emph{Upper panel:} $68\%$, $90\%$, and $99\%$ confidence contour levels of the ISM density and SRCUT break frequency derived from the spectrum of region $e$. \emph{Lower panel:} $68\%$, $90\%$, and $99\%$ confidence contour levels of the ISM density and the ISM ionization parameter derived from the spectrum of region $e$. The red curves are isochrones at 100 yr and 250 yr after the interaction with the shock front.}
\label{fig:densbreaktau}
\end{figure}

The upper panel of Fig. \ref{fig:densbreaktau} shows the confidence contours of the ISM density and the synchrotron cut-off frequency obtained from the spectrum of region $e$. In contrast to previous observations (\citealt{abd07}, \citealt{mbi09}), it is now possible to rule out the scenario where the ISM component is not present and the cut-off frequency is high ($\nu_{break}>3\times10^{15}$ Hz).
The lower panel of Fig. \ref{fig:densbreaktau} shows the confidence contours of the ISM density and the ISM ionization parameter in region $e$. Since both these values depend on the plasma emission measure, it is possible to estimate the time elapsed after the shock impact. The figure includes isochrones in the ($n_{\rm ISM}$,~$\tau_{\rm ISM}$) space at 100 yr and 250 yr. 

\subsection{Spatial distribution of the plasma properties}

If efficient hadron acceleration is at work, we expect to observe a higher shock compression ratio and a lower shock temperature in regions closer to the nonthermal limbs. From the modeling of X-ray spectra, we can derive the electron temperature. However, it is still unclear how to relate $T_e$ with the shock temperature \citep{vin12}. We therefore consider as the most reliable indicators of hadron acceleration the variations in the ISM postshock density that can be directly derived through our spectral analysis (see Sect. \ref{DP}). To quantify the possible shock modification along the rim, we analyzed the spectra extracted from the partially overlapping regions [$a$, $b$, $c$, $d$], and [$f$, $g$, $h$] (shown in Fig. \ref{fig:1006}) that are closer than $e$ to the nonthermal limbs. 

We first verified that we do not get a significant reduction of the $\chi^2$ by leaving the chemical abundances in regions $a,~b,~c,~d$ as free parameters in the fitting. Therefore, we fixed the O, Ne, Mg, and Si abundances to the values derived in region $e$, where the contribution of thermal emission is maximum (in the other regions, the nonthermal continuum is higher). In regions $f$ and $g$, we found that O and Ne are instead significantly different from those in region $e$. For example, in region $g$ we get $\chi^2=588.7$ (507 d.~o.~f.) by letting O and Ne free and $\chi^2=636.7$ (509 d.~o.~f.) with O and Ne tied to the values of region $e$. According to the F-test, the probability that the improvement of the fit is not significant is $<2.5\times 10^{-9}$. The best-fit values of O and Ne in regions $e,~f,~g,~h$ are shown in Table 1. In region $h$, the abundances were tied to those of region $g$, because they were all consistent within $0.5~\sigma$.
We then verified that the ejecta temperature and ionization parameter do not vary significantly across the regions. Therefore, we fixed them to their average values ($kT_{ejecta}=0.397$ keV and $\tau=5.29\times10^{10}$ s cm$^{-3}$, respectively). 

In all the regions, we detected the ISM component. The best-fit parameters for all eight spectral regions are given in Table \ref{tab:res}.

\begin{center}
\begin{table*}[htbp]
\begin{center}
\caption{Best-fit parameters}
\resizebox{\textwidth}{!} {
\begin{tabular}{lccccccccc} 
\hline\hline

 Parameter                       &    Region $a$         &      Region $b$       &   Region $c$          &     Region $d$        &       Region $e$      &       Region $f$      &   Region $g$          &    Region $h$   &  \\ \hline
 $n_{\rm ISM}$ (10$^{-1}$cm$^{-3}$)  & $1.99^{+0.4}_{-0.17}$ & $1.91^{+0.08}_{-0.16}$&$1.83^{+0.09}_{-0.10}$ & $1.63^{+0.08}_{-0.16}$& $1.44^{+0.10}_{-0.11}$&  $1.47\pm0.12$        & $1.92^{+0.09}_{-0.2}$ & $1.94^{+0.2}_{-0.17}$  \\ 
 $kT_{\rm ISM}$ (keV)                & $1.36^{+0.11}_{-0.16}$& $1.36^{+0.08}_{-0.09}$&$1.37^{+0.12}_{-0.09}$ & $1.36^{+0.18}_{-0.10}$& $1.36^{+0.2}_{-0.11}$ &  $1.41\pm0.07$        & $1.7^{+0.4}_{-0.3}$   & $1.9\pm0.2$       \\
$\tau_{\rm ISM}$ ($10^8$ s cm$^{-3}$)&   $4.4\pm0.6$         &   $5.5^{+0.5}_{-0.3}$ &  $5.7^{+0.4}_{-1.2}$  &  $6.0^{+0.6}_{-0.4}$  & $7.1^{+0.9}_{-0.7}$   &  $5.7\pm0.5$          &      $2.0^{+0.5}$     & $2.4\pm0.4$     \\
$n_{ejecta}$ (10$^{-1}$cm$^{-3}$)&$6.82^{+0.09}_{-0.12}$ & $6.52^{+0.07}_{-0.08}$&$6.31^{+0.11}_{-0.06}$ & $5.59^{+0.09}_{-0.05}$&     $4.95\pm0.11$    & $5.21^{+0.11}_{-0.14}$&     $8.2\pm0.2$       &  $9.3^{+1}_{-0.2}$  \\ 
          O                      &      $4.0$ (frozen)   &   $4.0$ (frozen)     &   $4.0$ (frozen)       &     $4.0$ (frozen)    &    $4.0\pm 0.2$	 & $4.4\pm0.2$           &   $4.5\pm0.2$         &  $4.5$ (frozen)   \\
          Ne                     &    $1.27$ (frozen)    &   $1.27$ (frozen)    &    $1.27$ (frozen)     &    $1.27$ (frozen)    &     $1.27\pm0.10$     & $1.50^{+0.14}_{-0.12}$&   $1.80\pm0.14$       &  $1.80$ (frozen) \\
          Mg                     &   $15.5$ (frozen)     &    $15.5$ (frozen)   &    $15.5$ (frozen)     &    $15.5$ (frozen)    & $15.5^{+1.2}_{-0.8}$  &  $15.5$ (frozen)      &  $15.5$ (frozen)      & $15.5$ (frozen)  \\
          Si                     &     $80$ (frozen)     &   $80$ (frozen)      &    $80$ (frozen)       &      $80$ (frozen     & $>75$ (frozen to 80)  &    $80$ (frozen)      &   $80$ (frozen)       & $80$ (frozen)   \\
  $\nu_{break}$ ($10^{15}$ Hz)  &$   6.9^{+0.5}_{-1.1}$  & $2.1^{+1.3}_{-1.1}$  & $2.3^{+0.7}_{-1.3}$    &     $1.0^{+1.4}$      &   $1.0^{+0.9}$        &     $1.0^{+1.0}$      &  $1.0^{+1.4}$         & $7\pm2$   \\
        $\chi^{2}/$d.~o.~f.      &      $962.1/710$      &      $1159.1/713$    &    $1235.1/720$        &       $1215.8/672$    &    $1113.9/657$       & $649.0/459$           &   $588.7/507$         & $1570.6/1160$   \\ 
\hline\hline   
\multicolumn{9}{l}{\footnotesize{All errors at $90\%$ confidence level. $kT_{ejecta}$ and $\tau_{ejecta}$ are frozen to $0.397$ keV and $5.29\times10^{10}$ s $cm^{-3}$, respectively.}} \\
\multicolumn{9}{l}{\footnotesize{The densities are estimated from the emission measure by calculating the volume of the emitting plasma and by}}\\
\multicolumn{9}{l}{\footnotesize{assuming pressure equilibrium between ejecta and ISM.}} \\
\label{tab:res}
\end{tabular}
}
\end{center}
\end{table*}
 \end{center}

\begin{figure}[htb!]
 \centerline{\hbox{     
     \psfig{figure=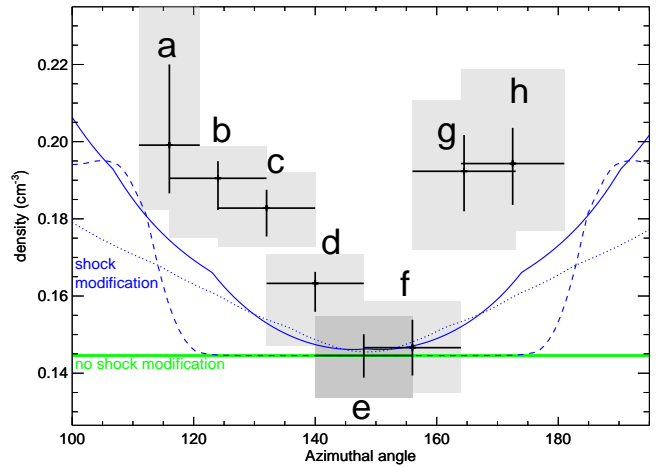,angle=90,width=\columnwidth}}}
 
\caption{Azimuthal profile of the postshock ISM density (see regions in Fig. 1). Error bars are shown at the $68\%$ (crosses) and $90\%$ (shaded areas) confidence levels. The blue curves show the profiles predicted by \citet{vbk03} with $\delta B/B=1$ (dot) and 0.8 (dashed), and derived by the radio flux (solid curve, see text) by assuming the relationship between the injection efficiency and shock compression ratio adopted by \citet{fdb10}. The green line indicates the expected trend in the case of no shock modification.}
\label{fig:dens}
\end{figure}

While we did not observe any clear trend in the $T_e$ azimuthal profile, we found that the postshock density significantly increases in regions where particle acceleration is more efficient (Table \ref{tab:res} and Fig. \ref{fig:dens}). Considering that the preshock density is expected to be fairly uniform \citep{dgg02}, the density profile in Fig. \ref{fig:dens} shows that \emph{the shock compression ratio is higher near the nonthermal limbs}. This result looks consistent with the predictions of the shock modification theory. 

It has been shown that where particle acceleration is efficient, models that do not account self-consistently for the acceleration process (as VPSHOCK) can provide biased results \citep{psr10}. We did not analyze spectra extracted from the nonthermal limbs, and we do not expect that particle acceleration is at its maximum efficiency in regions $a-h$. However, some effects may be present in the regions closer to the nonthermal limbs, where ``moderate'' acceleration is at work (see Sect. \ref{Conclusions}). Nevertheless, \citet{psr10} showed that, in conditions of moderate particle acceleration, models that do not account for particle acceleration provide inaccurate results below 1 keV. They also showed that the main problems are visible with high statistics (much higher than that of our observations) at energies corresponding to the O lines, which are not well modelled (this issue may somehow affect the estimation of the ionization parameter). We can reasonably conclude that our estimates of the density of 
the ISM 
component, whose contribution is higher at energies $>2$ keV (see Fig. \ref{fig:spec}), are reliable and that the density enhancement toward the nonthermal limbs is robust. A more detailed comparison between the available spectral models and models including the effects of the acceleration process is beyond the scope of this work, but is required to address this issue.

\begin{figure}[htb!]
 \centerline{\hbox{     
     \psfig{figure=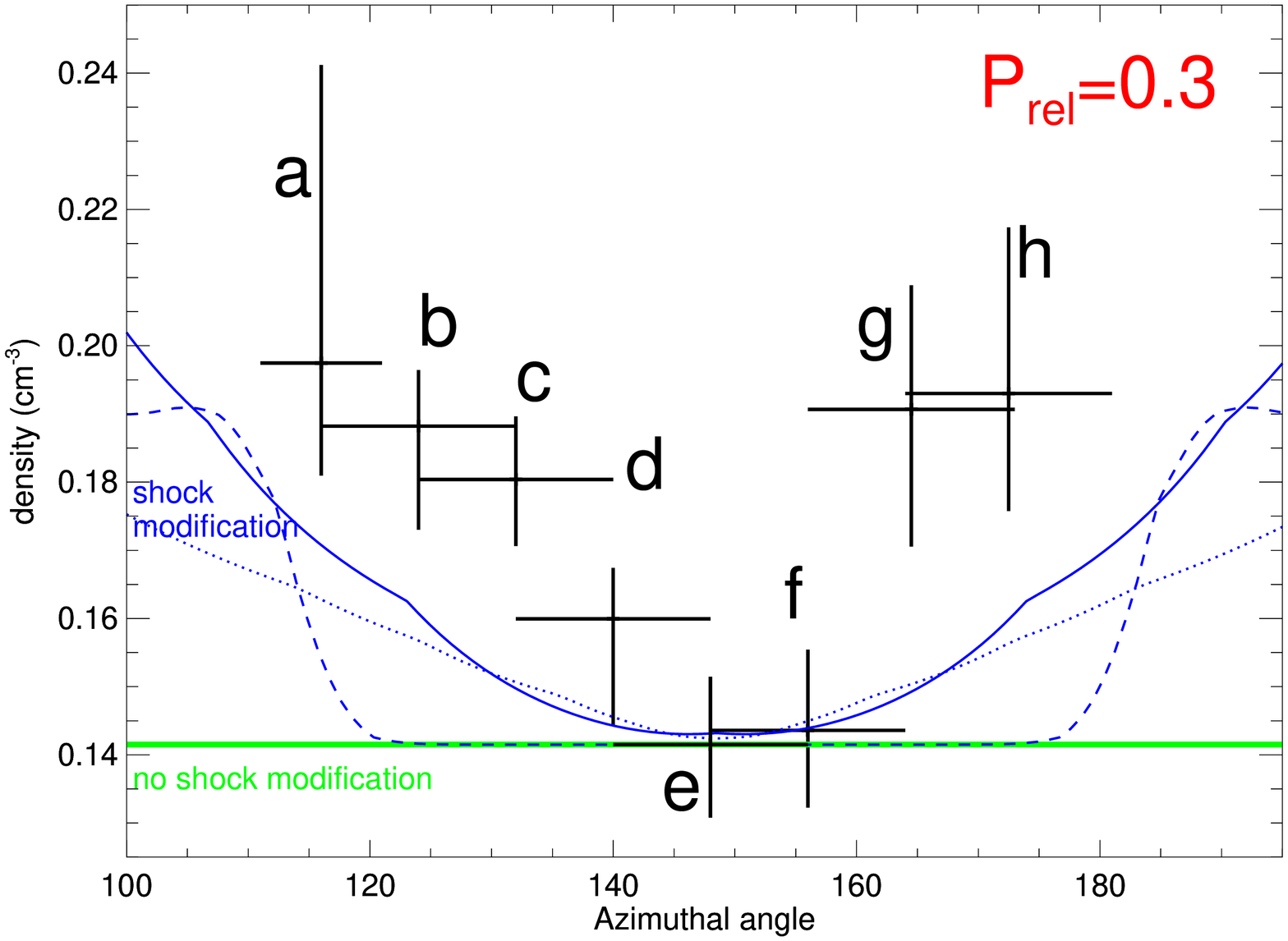,angle=90,width=\columnwidth}}}
      \centerline{\hbox{     
     \psfig{figure=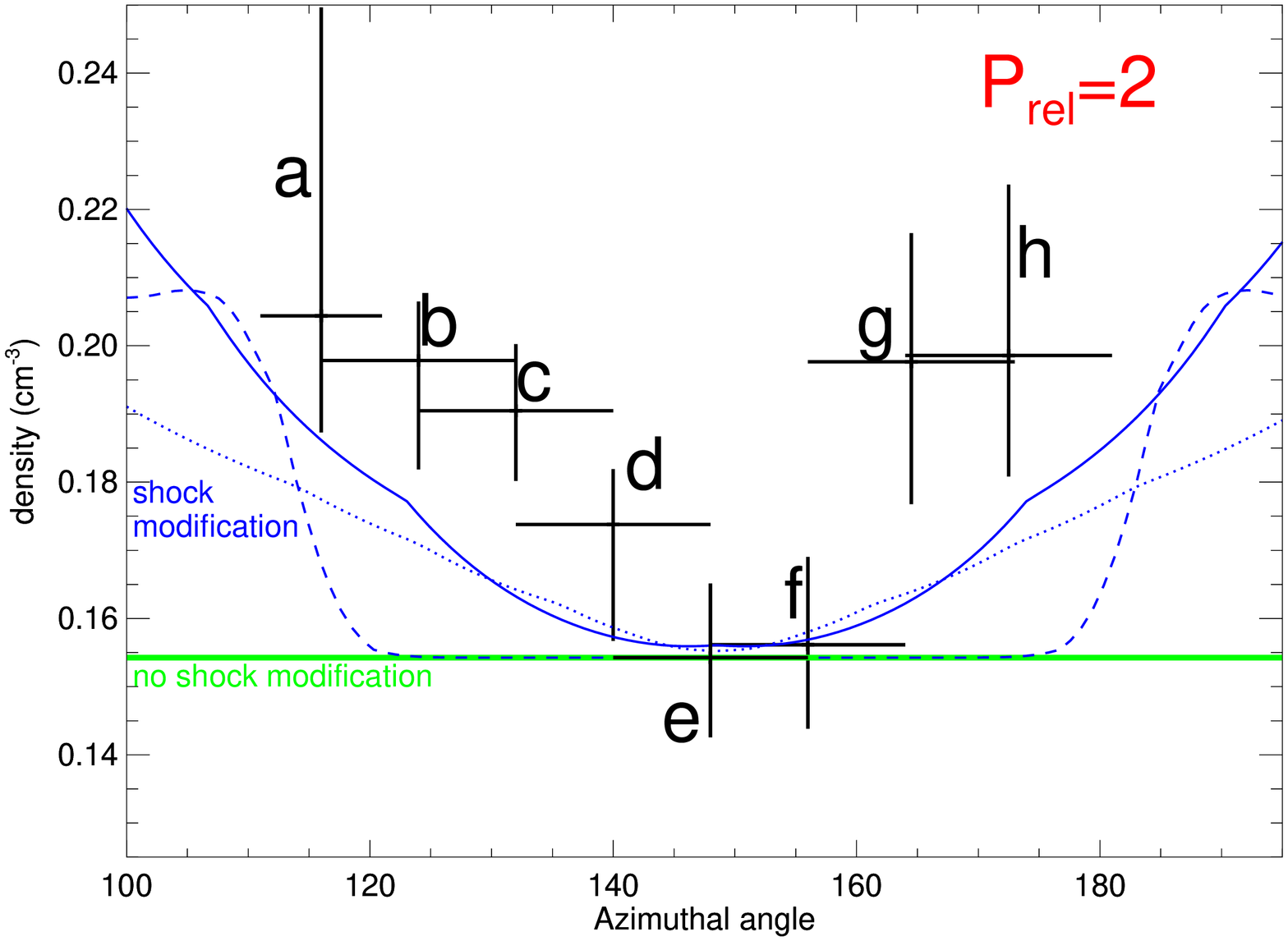,angle=90,width=\columnwidth}}}   
\caption{\emph{Upper panel}: ISM densities derived in regions $a-h$ of Fig. 1 by assuming that ratio of the ejecta to the ISM pressure is $P_{rel}=0.3$. Errors at $90\%$ confidence level. \emph{Lower panel:} Same as upper panel for $P_{rel}=2$.}
\label{fig:dens03-2}
\end{figure}

We point out that the values of the ISM density (derived by the ISM emission measure as explained in Sect. \ref{DP}) do not change significantly by relaxing the assumption of pressure equilibrium between ISM and ejecta. Indeed, the increase in the ISM density toward the nonthermal limbs shown in Fig. \ref{fig:dens} is also recovered by assuming a more realistic scenario, where the ejecta pressure is lower than the ISM pressure. Detailed 3-D hydrodynamic modeling of SN~1006 \citep{obm12} shows that the average ratio of the ejecta to the ISM pressure is $P_{ej}/P_{\rm ISM}=P_{rel}\sim0.3$. Therefore, as a further check, we derived the ISM densities as in \citet{bms99} by assuming $P_{rel}=0.3$ and obtained the results shown in the upper panel of Fig. \ref{fig:dens03-2}. The lower panel of the same figure shows that, even in the unrealistic scenario of $P_{rel}=2$ (i. e. the ejecta pressure is twice the ISM pressure), the ISM density near the nonthermal limbs is significantly higher than in the center of the 
thermal limb. 

We found that the filling factor of the ISM, $f_{ISM}$, in all the spectral regions is higher than 95\%. Hence, the volume fraction occupied by the ejecta is relatively small. This is in qualitative agreement with detailed magneto-hydrodynamic simulations of SN~1006 that show the presence of narrow fingers of ejecta close to the shock front \citep{obm12}.

Table 1 shows that the ejecta density is somehow correlated with the ISM density. This may suggest that effects of shock modification are present also at the reverse shock. However, while our estimates of the ISM density are robust, the ejecta densities may be less reliable, since they strongly depend on our estimates of the filling factors. The small relative variations of $f_{\rm ISM}$ in regions $a-h$ ($f_{\rm ISM}\sim0.95-0.98$), in fact, correspond to large relative variations of the ejecta filling factor ($f_{ejecta}\sim 0.02-0.05$) that strongly affect the estimates of the ejecta density. For example, by assuming that the filling factor of the ejecta is the same in all the regions and is equal to its average $<f_{ejecta}>=0.031$ (and $P_{rel}=1$), the ejecta density in region $a$ and $e$ are $n_{ej}\sim0.52$ cm$^{-3}$ and $n_{ej}\sim0.60$ cm$^{-3}$, respectively (to be compared with the values reported in Table \ref{tab:res}). On the other hand, we verified that the 
estimates of the ISM density are not affected by this issue.

Table 1 also shows the values of the cut-off frequency $\nu_{break}$, which increases toward the nonthermal limbs, as expected. With respect to \citet{mbi09} (where larger regions were adopted), our values are slightly smaller. This effect is due to the presence of the ISM component (not included in \citealt{mbi09}). Because of its extreme underionization ($\tau<7\times10^8$ s cm$^{-3}$), the ISM component is almost featureless, and in the archive observations where the statistics were not very high, it was possible to model its contribution by increasing the cut-off frequency of the SRCUT component.

The values of the ejecta ionization parameter in all regions is compatible with $\sim5\times10^{10}$ s cm$^{-3}$. These values are a factor $\sim2-3$ higher than those reported in \citet{mbi09}. We verified that these discrepancies are associated with the upgraded version of the VPSHOCK model (based on APED) that we are adopting here. Since the ejecta densities are $\sim0.6$ cm$^{-3}$, we derive that the plasma was shocked $t_{shock}\sim2500$ yr ago, i.~e., more than the age of SN~1006. This puzzling result relies on the assumption that the density remains constant over time. Lower values of $t_{shock}$ can be obtained by considering that the ejecta density decreases during the evolution of the remnant (because of the expansion). A lower value of $t_{shock}$ can also be obtained by assuming that the ejecta component consists of a ``pure ejecta" plasma, where the number of free electrons is much higher than the number of positive ions (because heavy elements are much more abundant than hydrogen). We verified 
that by increasing the chemical abundances by a factor of 1000 in the ejecta component and by reducing the emission measure $n_{e}n_{H} V$ by a similar factor (with respect to the values that can be derived from Table 1) we still obtain a good fit in region $e$. This indicates that a pure ejecta scenario is plausible.
In this case, the ejecta emission measure ($\sim1000n_{e}n_{H}V$) would be almost the same as that reported in Table 1, but the electron density would be higher by a factor $\sqrt{Z}$ (where $Z$ is the average number of free electrons per ion) and the time elapsed after the shock impact would be reduced by the same factor. 
We will investigate in detail this scenario in a forthcoming paper (Miceli et al., in preparation); we here only note that, even within this pure ejecta model, the normalization of the ISM component does $not$ vary significantly.

\section{Discussion and conclusions}
\label{Conclusions}

We have analyzed a set of deep \emph{XMM-Newton} observations of SN~1006 that have allowed us to address important issues.

We have detected X-ray emission from the shocked ISM in the southeastern rim of SN~1006. It was not possible to explore the conditions of the shocked ISM in the nonthermal limbs because the higher contribution of the synchrotron radiation did not allow us to obtain tight constraints. Nevertheless, in regions $a-h$, the high statistics of the new observations showed that the addition of a thermal, underionized component improves the quality of our spectral modeling. A further indication that the association of this additional component with the ISM is sound is that its contribution to the nitrogen line emission is significant. Nitrogen line emission cannot be associated with ejecta in remnants of Type Ia supernovae and should originate in the shocked ISM. As shown in Fig. \ref{fig:spec}, the contribution of the ejecta component dominates below 0.5 keV (i.~e., the band of the N lines). However, we found that the fit of our model to the spectra does not change significantly if we set the nitrogen abundance in 
the ejecta component to zero, while by putting N=0 in the ISM component, we obtained significant residuals around the nitrogen line. This indicates that in this band the contribution of the ISM component to the nitrogen emission lines is, as expected, stronger then the contribution of the ejecta component to the same line.

Figures \ref{fig:dens} and \ref{fig:dens03-2} clearly indicate that, under the assumption of constant preshock density, the shock compression ratio increases (up to $\sim5.5-6$) in regions where nonthermal emission is stronger. This is qualitatively consistent with the expectations from modified shock models.
In particular, such a high shock compression ratio indicates that the fractional cosmic-ray pressure in the postshock region is $\sim30\%$ (\citealt{vyh10}) near the nonthermal limbs.

Though a detailed theoretical modeling of the expected azimuthal profile of the shock compression ratio in SN~1006 is not available in the literature, we compared the observed ISM density profile with that predicted by the shock modification model provided by \citet{vbk03} and shown as dashed$/$dotted blue curves in Fig. \ref{fig:dens}. The solid blue curve was instead derived by assuming that the injection efficiency\footnote{The injection efficiency is the fraction of particles injected in the acceleration process.}, $\eta$, is proportional to the radio flux divided by $B^{3/2}$ (and has its minimum at $\eta=5\times10^{5}$ in region $e$), and adopting the magnetic field model MF2 of \citet{pbm09}. The green line shows the constant compression-ratio scenario, which corresponds to no shock modification. For all the curves, the relationship between the injection efficiency and the shock compression factor has been obtained by following \citet{fdb10}.
The shock modification models predict an azimuthal trend similar to the measured one, though they cannot fit all the points.
In particular, the models shown in Fig. \ref{fig:dens} slightly underpredict the observed density variations. However, we point out that a complete, self-consistent model of hadron acceleration at oblique shocks is not available yet and that current models do not provide tight theoretical constraints, though they allow us to get qualitative indications. Moreover, the models available do not include a dependence of the compression ratio on the maximum energy of the accelerated hadrons which may be higher near the nonthermal limbs. The cut-off frequency of the synchrotron emission from ultrarelativistic electrons is, in fact, much higher in the limbs than in the center and elsewhere in the rim \citep{rbd04,mbi09}. This indicates that electrons are accelerated to higher energies in the limbs and suggests that protons may also be accelerated to higher energies. 

In conclusion, we have detected the contribution of the shocked ISM to the X-ray emission in the southestern rim of SN~1006. We also found an azimuthal trend of the ISM postshock density that suggests that cosmic ray acceleration is at work at the shock front of SN~1006 and modifies the shock compression ratio. 
The results we have obtained in the southeastern rim of SN 1006 therefore indicate the presence of shock modification by particle acceleration. 
A definitive proof will require more data at higher spatial resolution. These data are necessary to extract spectra from narrow regions between the shock front and the contact discontinuity in order to better isolate the contribution of the shocked ISM and to confirm the effect that we have detected in this work.

\begin{acknowledgements}
 The authors thank the anonymous referee for their useful comments. This paper was partially funded by the ASI-INAF contract I$/$009$/$10$/$0. AD and GM acknowledge support from the CNES. The authors thank Jean Ballet for interesting discussions and suggestions.
 \end{acknowledgements}

\bibliographystyle{aa}


\newpage

\appendix

\section{Volume calculation}

We here describe the procedure we developed to calculate the volume of the emitting plasma in the spectral regions shown in Fig.~1. 

\begin{figure}[htb!]
 \centerline{\hbox{     
     \psfig{figure=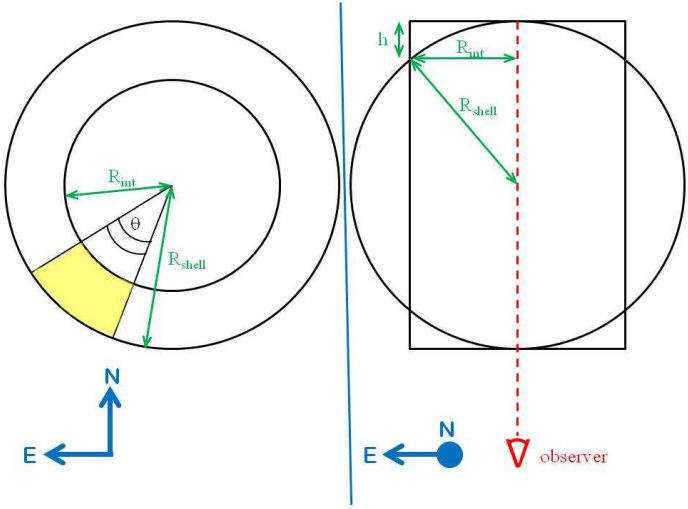,width=\columnwidth}}}
\caption{Schematic representation of SN~1006 (sphere with radius $R_{shell}$) and of the spectral regions shown in Fig. 1 (yellow area)}
\label{fig:vol}
\end{figure}

Let us assume that the SN 1006 shell is a sphere with radius $R_{shell}$ and let us consider a cylinder with radius $R_{int}$ and height $2R_{shell}$. In agreement with \cite{chr08}, we consider as the center of the shell the point with coordinates $\alpha _{J2000}=15^{h} 02^{m} 56.8^{s}$,$\delta _{J2000}) = -41^{\circ }56^{\prime }56.6^{\prime \prime })$ (see Fig. 1). We also assume a distance of 2.2 kpc. Let $R_{shell}$, $R_{int}$, and $\theta$ be the outer radius, inner radius, and opening angle of our spectral regions, respectively.  Figure \ref{fig:vol} shows the projection of the sphere and the cylinder in the plane of the sky (left panel) and in the plane identified by the line of sight and the east direction (right panel). Since the volume of the spherical cap with height $h=R_{shell}-(R^{2}_{shell}-R^{2}_{int})^{1/2}$ is $V_{cap}=\pi h/6(3R^{2}_{int}+h^2)$, the volume of the intersection between the cylinder and the sphere is 
\begin{equation}
V_{c-s} = \pi R^{2}_{int}\times2R_{shell} - 2(\pi R^{2}_{int} h -V_{cap})
\end{equation}

The volume of the portion of the sphere outside of the cylinder is $V_{0}=4/3\pi R^{3}_{shell}-V_{c-s}$ and the volume of our spectral region is $V_{reg}=\theta/2\pi V_0$ which depends only on the outer radius, inner radius, and opening angle of the region. 

For each region, we then measured $R_{shell}$ (SN~1006 is not a perfect circle in the sky and small azimuthal variations are present), $R_{int}$, and $\theta$ for each region, thus obtaining $V_{reg-a}=2.3\times10^{56}$ cm$^3$, $V_{reg-b}=2.7\times10^{56}$ cm$^3$, $V_{reg-c}=3.4\times10^{56}$ cm$^3$, $V_{reg-d}=3.0\times10^{56}$ cm$^3$, and $V_{reg-e}=3.2\times10^{56}$ cm$^3$.

\newpage
\section{Analysis of the pn spectra}
\begin{figure}[htb!]
 \centerline{\hbox{     
     \psfig{figure=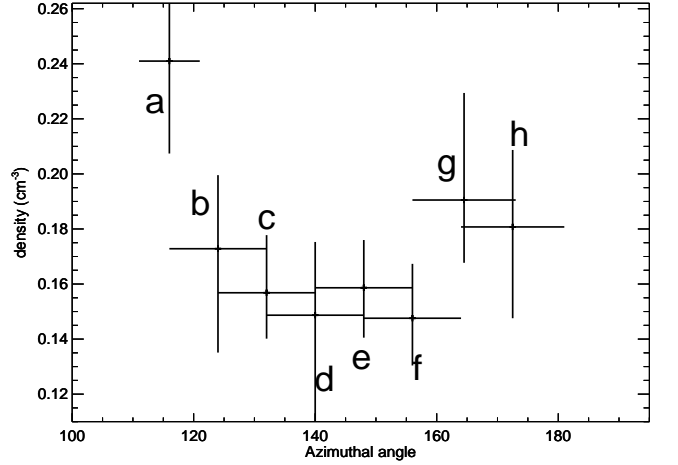,angle=90,width=\columnwidth}}}
\caption{ISM densities derived in regions $a-h$ of Fig. 1 by fitting the pn spectra only in the $0.6-7$ keV band. Errors at $90\%$ confidence level.}
\label{fig:pn}
\end{figure}

To further test the robustness of the density profile shown in Fig. \ref{fig:dens}, we also fitted the pn spectra only in the $0.6-7$ keV band. Alhough in this case the best-fit parameters are more sensitive to the background subtraction procedure (specially the ISM temperature) and error bars are larger, we still obtained an ISM density profile consistent with that of Fig. \ref{fig:dens}. Figure \ref{fig:pn} shows the density profile obtained by fitting the pn spectra only and by adopting method $1$ (see Sect. \ref{DP}) for the background subtraction.

\end{document}